\documentclass[french]{pfia}
\usepackage{amsthm}
\usepackage{amsmath}
\usepackage{algorithmfr}
\usepackage{algorithmicfr}
\usepackage{graphicx}
\usepackage{placeins}
\usepackage{latexsym}
\usepackage{amssymb}
\usepackage{multicol}
\usepackage{hyperref}
\usepackage{supertabular}
\usepackage{array,multirow,makecell}
\setcellgapes{1pt}
\makegapedcells
\newcolumntype{R}[1]{>{\raggedleft\arraybackslash }b{#1}}
\newcolumntype{L}[1]{>{\raggedright\arraybackslash }b{#1}}
\newcolumntype{C}[1]{>{\centering\arraybackslash }b{#1}}

\title{\textbf{Recommandation ontologique multicritère pour la métrologie}}

\author{Axel Mascaro, Christophe Rey\\[6pt]
	Université Clermont Auvergne, LIMOS\\[6pt]
	\{axel.mascaro,christophe.rey\}@uca.fr}

\date{}

\begin{document}
	
	\maketitle
	
	
	\begin{resume}
		La recommandation, le classement d'informations sont des processus d'aide à l'utilisateur qui lui propose la ou les meilleures réponses à sa requête. En l'absence d'une réponse exacte, lui offrir les propositions les plus proches possibles de son besoin est une amélioration de ce processus. Dans le contexte d'une plateforme de recherche d'information métrologique, nous discuterons du problème de classement par l'intermédiaire de la représentation des connaissances, avec une approche basée sur les Logiques de Descriptions, les ontologies et la comparaison multicritère. Nous proposerons un raisonnement permettant de comparer entre elles plusieurs propositions, à l'aide notamment de la différence et la découpe en plusieurs composantes d'une information. Nous présenterons une heuristique mettant en application ces résultats.
	\end{resume}
	
	\begin{motscles}
		Recommandation, classement, logique de description, Ontologie, comparaison multicritère. 
	\end{motscles}
	
	\begin{abstract}
		Matchmaking and information ranking are helping process for users, by offering them the best answers possible at their request. When there is no exact answer, giving them the closest proposition available is an efficient upgrade of that helping process. With a reasearch platform on metrology as a framework, we will discuss about ranking with knowledge representation, with an approach based on Description Logic, ontologies and multricriteria comparison. We present a reasonning to compare each proposition with the other, with semantic and syntaxic difference, by troncating the information in distinct component. 
	\end{abstract}
	
	
	
	\section{Introduction}
	\label{intro}
	
	STAM\footnote{\url{https://limos.fr/news_project/118}} est un projet d'une plateforme multi-usage dans le domaine de la métrologique. Elle permettra à n'importe quel industriel de standardiser une description d'une mesure, la comparer avec d'autres existantes et obtenir des valeurs et une incertitude révisée en fonction de son contexte de mesure. Le projet cherche aussi à créer une communauté autour de la métrologie et de fournir une plateforme de documentation sur le sujet. C'est sur cette dernière partie que ce travail porte. 
	
	
	La plateforme en cours propose déjà un système de recherche de ressources. Une ressource représente une information ou un fichier stocké sur la plateforme. Elle est identifiée par des caractéristiques définies dans une ontologie de la métrologie et qui sont utilisées lors des recherches. Les facettes sont des annotations qui font partie de ces caractéristiques et elles sont communes à plusieurs ressources. Par la sélection d'une facette, l'utilisateur va pouvoir restreindre les résultats de sa recherche aux ressources annotées par la facette choisie. Par exemple, la facette "Instrument" permet de trier les ressources obtenues en réponse à une recherche en fonction des instruments de mesure évoqués dans les ressources. Il est possible pour l'utilisateur de choisir plusieurs facettes et d'utiliser la recherche par mots-clés en même temps.
	
	Nous proposons de compléter cette recherche par un processus de recommandation de ressources basée sur une recherche approximative par rapport à la sémantique de leurs descriptions. On cherche alors à offrir, en plus des résultats exacts lors de recherche par facette, les résultats approximatifs les plus pertinents : les ressources les plus proches sémantiquement des facettes et mots-clés de la requête. Nous proposons ainsi d'envisager la recherche approximative de ressources comme un raisonnement de plus proche mise en correspondance sémantique (matchmaking sémantique) de descriptions de ressources avec une requête utilisateur, basé sur une ontologie du domaine de la métrologie et piloté par une approche de comparaison multicritère.
	
	En section \ref{LD} nous faisons des rappels concernant les logiques de description, formalisme de représentation des connaissances et de raisonnement utilisé ici pour définir la proximité sémantique d'une ressource avec une requête. A la section \ref{recherche}, nous proposons notre raisonnement. Plus précisément, en section \ref{composante}, nous définissons la notion de composante nécessaire à la comparaison des requêtes et ressources. Puis, en section \ref{semantic}, nous formalisons la notion de classement sémantique dans cadre des logiques de description permettant le tri des ressources au niveau de chaque composante. En section \ref{rankingMulti}, nous proposons une première heuristique multicritère synthétisant le classement sémantique sur l'ensemble des composantes. En section \ref{panorama}, nous évoquons quelques travaux existants de matchmaking sémantique et nous situons notre approche par rapport à eux. Enfin, nous concluons en section \ref{conclu}.

	\section{Logiques de descriptions}
	\label{LD}
	
	Les logiques de description (LD) est un formalisme de représentation des connaissances et de raisonnement sur celles-ci. Plus précisément, elles constituent une famille de sous-langages de la logique du premier ordre munies de la même sémantique basée sur la théorie des modèles, mais aussi d'une syntaxe particulière basée sur les notions de concept, de rôle et d'individu. 
	On considère que l'on a un ensemble d'éléments appelés domaine. Un concept est identifié par un nom comme $Instrument$ ou $Material$. Un concept définit un ensemble d'éléments du domaine, comme le ferait une classe en langage objet. Un rôle est identifié par un nom comme $hasMaterial$ ou $hasInstrument$. Un rôle définit un ensemble de couples d'éléments du domaine. Par exemple, le rôle $hasMaterial$ relie ici un instrument avec son matériau constituant. Il définit un ensemble de couples d'éléments dont le premier est un instrument et le second est le matériau constituant cet instrument. Un individu est identifié par un nom comme $RegleBois2$ ou $Acier1$. C'est un élément du domaine. 
	
	Plus formellement, la sémantique des connaissances décrites par les LD est définie par la notion d'interprétation. Une interprétation est une paire \ensuremath{\mathcal{I}} = (\ensuremath{\Delta^{\mathcal{I}}}, .\ensuremath{^{\mathcal{I}}}), où \ensuremath{\Delta^{\mathcal{I}}} est le domaine qui regroupe l'ensemble des individus étudiés et .\ensuremath{^{\mathcal{I}}} la fonction d'interprétation qui relie tous les concepts à une partie de \ensuremath{\Delta^{\mathcal{I}}}, et chaque rôle à une partie de \ensuremath{\Delta^{\mathcal{I}}} \ensuremath{\times} \ensuremath{\Delta^{\mathcal{I}}}. 
	
	Ces trois éléments (concepts, rôles et individus) peuvent se combiner par l'intermédiaire de constructeurs pour former des descriptions de concepts et de rôles plus complexes qu'un simple nom. Les descriptions élémentaires sont appelées concepts et rôles atomiques (ce ne sont que des noms). Chaque logique de description est définie par sa propre combinaison de constructeurs. 
	On peut citer :
	\begin{itemize}
		\item la conjonction de concepts notée ~\ensuremath{C \sqcap D} qui permet de créer l'intersection des descriptions de concept $C$ et $D$;
		\item la quantification existentielle (resp. universelle), notée $\exists R.C$ (resp. $\forall R.C$) avec $R$ un nom de rôle et $C$ une description de concept, qui permet de définir l'ensemble des individus liés par $R$ à au moins un individu lui-même appartenant à $C$ (resp. liés par $R$ uniquement à des individus de $C$);
		\item la négation, notée $\neg C$ permettant de définir la description de concept regroupant tous les individus qui ne sont pas dans $C$.
	\end{itemize} 
	Par exemple, $Regle$ ~\ensuremath{\sqcap} \ensuremath{\exists}$hasMaterial.Steel$ décrit l'ensemble des règles dont au moins un des matériaux le constituant est l'acier. Par ailleurs $Regle$ ~\ensuremath{\sqcap} \ensuremath{\forall}$hasMaterial.Steel$ décrit l'ensemble des règles dont le matériau les constituant est l'acier.
	
	Le tableau \ref{tab} regroupe la syntaxe et la sémantique des constructeurs de deux logiques de description : ${\cal ALU}$ et ${\cal EL}^+$. ${\cal EL}^+$ est une logique de description introduite dans ce travail et qui est égale à la logique ${\cal EL}^{++}$ étudiée dans \cite{PushingEL} sans le constructeur des domaines concrets. Nous justifions plus bas l'intérêt de ce langage.
	
	\begin{table}[!htbp] 
		\small{
			\begin{tabular}{ |p{1.5cm}|p{1cm}|p{2.40cm}|p{0.6cm}|p{0.6cm}| }
				\hline
				Construct. & Syntaxe & Semantique &   \ensuremath{\mathcal{ALU}} &  ${\cal EL}^{++}_{\setminus {\cal D}}$\ \\
				\hline \hline
				concept atomique & $P$ & $P^{\cal I} \subseteq \Delta^{\cal I}$ &   X & X\\ \hline
				rôle atomique & $R$ & $R^{\cal I} \subseteq \Delta^{\cal I} \times \Delta^{\cal I}$ &   X & X\\ \hline
				top & $\top$ & $\Delta^{\cal I}$ &  X & X\\ \hline
				bottom & $\bot$ & $\emptyset$ &  X & X\\ \hline
				nominal & $\{a\}$ & $\{a^{\cal I}\}$ & & X \\ \hline
				conjonction & $C \sqcap D$ & $C^{\cal I} \cap D^{\cal I}$ &   X & X\\ \hline
				négation atomique& $\neg P$ & $\Delta^{\cal I} \setminus P^{\cal I}$ &  X &\\ \hline
				quantif. universelle & $\forall R.C$ & $\{x \in \Delta^{\cal I} |$
				$\forall y:$ $(x,y) \in R^{\cal I}$ $\rightarrow y \in C^{\cal I} \}$ &  X & \\ \hline
				quantif. existentielle & 	$	\exists R.C $ & $\{x \in \Delta^{\cal I} |$
				~$\exists y \in \Delta^{\mathcal{I}} :$ $(x,y) \in r^\mathcal{I}~ \wedge y~ \in ~C^\mathcal{I} \}$ & X $^1$  & X\\ \hline
				disjonction  & $	C ~\ensuremath{\sqcup} D $ &  $C^{\cal I} \cup D^{\cal I}$ &   X & \\ \hline
				composition de rôles & $R \circ S$ & $\{(x,z)\in (\Delta^{\cal I})^2 ~|$ $~ \exists y \in \Delta^{\cal I} : (x,y)\in R^{\cal I} \wedge (y,z)\in S^{\cal I}\}$ & & X \\ \hline
			\end{tabular}
			\caption{Syntaxe et sémantique des constructeurs définissant les LD ${\cal ALU}$ et ${\cal EL}^{++}_{\setminus {\cal D}}$, avec $P$ un concept atomique, $C$ et $D$ des descriptions de concepts, $R$ et $S$ des rôles atomiques et $a$ un nom d'individu. Note : $^1$ Dans ${\cal ALU}$ on ne peut construire que $\exists R.\top$  et non $\exists R.C$.}
			\label{tab}
		}
	\vspace{-0.4cm}
	\end{table}

	Les descriptions de concepts peuvent être reliées entre elles par des axiomes, notamment de subsomption. Les axiomes de subsomption (appelés GCI pour global concept inclusion), notés $C \sqsubseteq D$, où $C$ et $D$ sont des descriptions de concept spécifiant que tous les individus de $C$ sont aussi des individus de $D$. 
	On peut diviser les instruments en deux catégories disjointes en fonction de leur mode de lecture (analogique ou numérique) grâce aux axiomes suivants :
	~\\
	$Instrument$ ~\ensuremath{\sqsubseteq} $analogicInstrument$\ensuremath{\sqcup}\\ $\text{~~~~~~~~~~~~~~~~~~~~~~~~~~~~}numericInstrument$\\
	$analogicInstrument$ ~\ensuremath{\sqsubseteq}  ~\ensuremath{\neg} $numericInstrument$
	
	Pour les rôles, on peut exprimer des axiomes d'inclusion de rôles du type $R \sqsubseteq S$, avec $S$ un rôle atomique et $R$ un rôle atomique ou une composition de rôles atomiques. 
	
	En plus des axiomes, on peut exprimer des assertions de concept (resp. de rôle), pour postuler que tel individu (resp. tel couple d'individus) est dans l'interprétation de telle description de concept (resp. de tel rôle). Les assertions $Ruler(rul1)$, $Steel(steel2\%)$ et $hasMaterial(rul1, steel2\%)$ décrivent que $rul1$ est une règle, fait d'un acier particulier $steel2\%$.
	
	Un ensemble d'axiomes de concepts est appelé Terminological Box (TBox). Un ensemble d'axiomes de rôles est appelé Role Box (RBox). Une Constraint Box (CBox) est l'ensemble d'une TBox et d'une RBox. Un ensemble d'assertions est appelé Assertionnal Box (ABox). Dans la suite, on pourra utiliser le terme d'ontologie pour parler d'une CBox, ou d'une CBox et d'une ABox. La table \ref{tabEL+} donne la liste des axiomes et assertions disponibles dans ${\cal EL}^{++}_{\setminus {\cal D}}$.
	%
	\begin{table}[!htbp] 
		\small{
			\begin{tabular}{ |p{3.6cm}|p{2.4cm}|p{0.9cm}| }
				\hline
				Axiome et assertions& Syntaxe 
				
				et Sémantique &  ${\cal EL}^{++}_{\setminus {\cal D}}$\ \\
				\hline
				General Concept Inclusion  & $ C \ensuremath{\sqsubseteq} D$ 
				
				  $C^{\cal I} \subseteq D^{\cal I}$ &   X\\ \hline
				Inclusion de rôle  & $	R_1 \circ \ldots \circ R_n \ensuremath{\sqsubseteq} R $ 
				
				 $ R_1^\mathcal{I} \circ \ldots \circ R_n^\mathcal{I} \subseteq r^\mathcal{I} $ &  X\\ \hline
				Assertion de concept & $C(a)$ 
				
				 $a^{\cal I}\in C^{\cal I}$ & X \\ \hline
				Assertion de rôle & $R(a,b)$ 
				
				 $(a^{\cal I},b^{\cal I}) \in R^{\cal I}$ & X \\ \hline
			\end{tabular}
			\caption{Syntaxe et sémantique des axiomes et assertions définissant la LD ${\cal EL}^{++}_{\setminus {\cal D}}$, $C$ et $D$ des ${\cal EL}^{++}_{\setminus {\cal D}}$-descriptions de concepts, $R$, $R_1$, $\ldots$, $R_n$ des rôles atomiques et $a$ et $b$ des noms d'individus.}
			\label{tabEL+}
		}
	\vspace{-0.3cm}
	\end{table}

	\newtheorem{exe}{Exemple}

	\begin{exe}[Exemple de CBox en métrologie]
		\label{excours}
		La CBox ci-dessous décrit un univers métrologique composé d'instruments et de mesures. Une mesure est décrite par une unité et une dimension, un instrument par son materiau constituant, son type et son mode de lecture. Le reste des axiomes permet de créer une hiérarchie des concepts. On spécifie ainsi que le fer est un métal, et que le métal est un matériau (et donc par raisonnement que le fer est un matériau). Nous utiliserons cette ontologie dans les exemples qui suivront et elle permettra de décrire des ressources parlant de mesure et de l'instrument utilisé pour celle-ci.
		
		\begin{supertabular}{R{2.3cm}C{0cm}p{3cm}}
			$Measure$	& $\sqsubseteq$ &$ \exists hasUnit.Unit \sqcap$ \\
			& & $\exists hasDimension.Dimension$ \\
			$Instrument$	& $\sqsubseteq$ &  $\exists hasMaterial.Material  \sqcap$\\
			& & $\exists hasInstrumentType.$ $InstrumentType \sqcap$\\
			& & $ \exists hasReadingMode.$ $ReadingMode$ \\
			$	Metal$	& $\sqsubseteq$ & $Material$ \\
			$	Steel$	& $\sqsubseteq$ & $Metal$ \\
			$	Iron$	& $\sqsubseteq$ &$ Metal$ \\
			$	Wood$	& $\sqsubseteq$ & $Material$ \\
			$	Oak $ & $\sqsubseteq$ & $ Wood$ \\
			$Analogic$  & $\sqsubseteq$ & $ ReadingMode$ \\
			$Numeric$  & $\sqsubseteq$ &  $ReadingMode$ \\
			$Length $  & $\sqsubseteq$ & $ Dimension $\\
			$Centimeter $  & $\sqsubseteq$ & $ Unit$ \\
			$Ruler$   & $\sqsubseteq$ &  $InstrumentType$\\
			$Calliper$   & $\sqsubseteq$ & $ InstrumentType$ \\
		\end{supertabular}
	\end{exe}
\vspace{-0.1cm}	
	Pour faciliter la présentation des exemples, les termes de métrologie $Material$, $Dimension$, $InstrumentType$ et $ReadingMode$ seront abrégées en $Mat$, $Dim$, $IT$ et $RM$. 
	
	On utilise deux raisonnements de base avec les LD, la satisfaisabilité et la subsomption. Le premier vérifie qu'une interprétation \ensuremath{\mathcal{I}} satisfait l'ensemble des axiomes d'une CBox et des assertions d'une ABox. Le second consiste à vérifier que pour toute interprétation \ensuremath{\mathcal{I}}, on a :
	$C \ensuremath{\sqsubseteq} D \ensuremath{\Leftrightarrow} C\ensuremath{^{\mathcal{I}}} \ensuremath{\subseteq} D\ensuremath{^{\mathcal{I}}}$. 
	On lit ainsi "$C$ est subsumé par $D$". Si $C$ est subsumé par $D$ et $D$ est subsumé par $C$, alors $C$ et $D$ sont équivalents, et on note $C\equiv D$. Si $C$ est subsumé par $D$ mais n'est pas equivalent à $D$, alors on note $C\sqsubset D$. On pourra aussi utiliser les notations inverses $\sqsupseteq$ et $\sqsupset$. Dans cet article, on raisonnera toujours par rapport à une CBox CB non vide et une ABox AB (potentiellement vide).

	Plus une LD possède de constructeurs, plus elle est expressive et se voit capable de modéliser finement le domaine qu'elle doit représenter. Cela s'accompagne néanmoins d'une complexité croissante lors de la résolution des raisonnements de base. Une difficulté de l'utilisation des LD réside donc dans leur choix, pour garantir un maximum d'expressivité tout en restant performant.
	
	Dans ce travail, nous avons décidé d'utiliser la LD ${\cal EL}^{++}_{\setminus {\cal D}}$, qui est égale à ${\cal EL}^{++}$ \cite{PushingEL} privée des domaines concrets. En effet, il a été prouvé dans \cite{PushingEL}, que ${\cal EL}^{++}$ conserve un temps polynomial pour la détermination de la subsomption et de la satisfiabilité (sauf pour certains domaines concrets particuliers) en présence d'une CBox et d'une ABox, tout en ayant un pouvoir expressif conséquent, contrairement à d'autres LD basées sur l'utilisation de la quantification existentielle. Par ailleurs,  ${\cal EL}^{++}$ est également munie d'une forme normale définie dans \cite{PushingEL} ce qui en facilite l'utilisation.
	
	 ${\cal EL}^{++}$ est connue pour être suffisamment expressive pour un grand nombre d'applications, ce que nous avons vérifié dans la construction de l'ontologie de la métrologie pour le projet STAM. N'ayant pas eu besoin pour le moment des domaines concrets, nous limitons notre étude à ${\cal EL}^{++}_{\setminus {\cal D}}$. Enfin, un dernier atout de cette LD est qu'elle est proche du profil ${\cal EL}$ de OWL2, le langage standard pour la modélisation d'ontologie sur le web, ce qui rend possible l'usage des outils associés existants (éditeurs et raisonneurs).
	
	\newpage
	\section{Recherche des ressources proches}
	\label{recherche}
	Nous détaillons dans cette section notre approche pour trouver les ressources les plus proches d'une requête donnée. Nous commençons en section \ref{composante} par expliquer la structure en composantes des descriptions des ressources et de la requête. En section \ref{semantic}, nous montrons comment classer les ressources par rapport à chaque composante avec un raisonnement en LD. En section \ref{rankingMulti}, nous proposons un classement des ressources qui intègre toutes les composantes par une approche multicritère.
	
	\subsection{Recherche avec composante}
	\label{composante}
	
	Nous postulons que les requêtes et ressources sont décrites selon plusieurs dimensions que l'on appelle composantes. 
	
	\begin{exe}
		\label{excompo}
		Un document pdf décrivant une mesure de diamètre avec un pied à coulisse numérique en acier peut être décrit par une composante spécifiant le type d'instrument (introduite par le rôle $hasInstrument$) et une composante spécifiant la mesure (introduite par le rôle $hasMeasure$):\\
		\begin{supertabular}{R{1.2cm}C{0cm}p{6cm}}
			$DocPDF$	& $\equiv$ & 
			$\exists hasInstrument.$ $(\exists hasIT.Calliper \sqcap$  
			$\exists hasRM.Numeric \sqcap$
			$\exists hasMat.Steel)$  
			
			$\sqcap$ 
			
			$\exists hasMeasure.$ 
			$(\exists hasUnit.cm \sqcap$ 
			$\exists hasDim.length)$\\
		\end{supertabular}
	\end{exe}
\vspace{-0.1cm}
	Cette structure de description offre l'avantage de diviser la comparaison globale d'une requête avec une ressource en plusieurs comparaisons (une par composante) plus simples à effectuer, sur le principe "diviser pour régner". 
	%
	%
	%
	%
	%
	La recherche consiste ainsi dans un premier temps à classer  les ressources les plus proches de la requête pour chaque composante, puis à agréger ces classements par composante pour en obtenir un unique global. Cette structure en composantes est obtenue, pour les ressources, grâce aux experts qui les décrivent, et pour les requêtes grâce à un processus qui n'est pas développé dans cet article permettant de générer automatiquement les descriptions par composante des requêtes utilisateur à partir des facettes et des mots-clés choisis par ces derniers. Par la suite, nous désignerons la requête de l'utilisateur par le terme "demande", et les ressources disponibles par le terme "offres". 
	
	Nous formalisons maintenant notre approche à base de composantes.
	\newtheorem{defi}{Definition}
	
	\begin{defi}[Composante]
		Etant donnée une ${\cal EL}^{++}_{\setminus {\cal D}}$-CBox CB et $E$ une description de concept de CB, la composante ${\cal C}_E$ est l'ensemble des ${\cal EL}^{++}_{\setminus {\cal D}}$-descriptions de concepts qui sont subsumées par $E$ dans CB (avant et après inférence). $E$ est appelée le top concept de ${\cal C}_E$.
	\end{defi}
	
	On suppose que, pour chaque composante ${\cal C}_E$ d'une CBox CB, il existe un rôle $R_{{\cal C}_E}$ dit "rôle de composante" associé à ${\cal C}_E$. On dira aussi que ${\cal C}_E$ est associée à $R_{{\cal C}_E}$. On suppose que le top concept $E$ de ${\cal C}_E$ est aussi la portée du rôle $R_{{\cal C}_E}$, ce qui signifie que l'axiome suivant devrait être vérifié dans CB : $\top \sqsubseteq \forall R_{{\cal C}_E}.E$. Le constructeur $\forall$ et le constructeur de portée pour un rôle n'étant pas disponibles dans ${\cal EL}^{++}_{\setminus {\cal D}}$, on suppose que toute description contenant $\exists R_{{\cal C}_E}.D$ implique une description $D$ qui appartient à ${\cal C}_E$. 
	
	\begin{exe}
		\label{excours2}
		Avec la CBox de l'exemple \ref{excours}, on définit deux composantes : Instrument et son top concept $Instrument$, et Measure avec son top concept $Measure$. On suppose l'existence dans la CBox deux rôles de composante associés $hasInstrument$ et $hasMeasure$ (cf exemple \ref{excompo}). 
	\end{exe}
	
	Par commodité de langage et quand le contexte sera clair, on pourra employer le terme composante soit dans le sens exact de sa définition, soit pour évoquer son top concept, soit pour évoquer son rôle de composante associé.
	
	\begin{defi}[Offre (resp. demande)]
		Etant donnée une ${\cal EL}^{++}_{\setminus {\cal D}}$-CBox  CB et ${\cal C}_{E_1}$, ..., ${\cal C}_{E_n}$ les $n$ composantes de CB (données arbitrairement), avec $R_1$, ..., $R_n$ les rôles de composantes associés, une offre $O$ (resp. une demande $D$) est une ${\cal EL}^{++}_{\setminus {\cal D}}$-description de concept qui s'écrit $O \equiv \exists R_1.C_1 \sqcap \ldots \sqcap \exists R_n.C_n $ où chaque $C_i$ une ${\cal EL}^{++}_{\setminus {\cal D}}$-description de concept appartenant à ${\cal C}_{E_i}$, $\forall i \in \{1,\ldots,n\}$.
	\end{defi}
	
	Toute offre et demande admet dans sa description un terme pour chaque composante. Si la composante $R$ n'est pas utile pour la description, alors elle est fixée à $\exists R.\top$.
	
	\begin{exe}
		\label{excours3}
		
		En reprenant la CBox de l'exemple \ref{excours} avec les composantes de l'exemple \ref{excours2}, on peut imaginer les demandes $D1$ et $D2$ qui recherchent des ressources sur des pieds à coulisse numériques en acier, et des règles en bois graduées en cm :
		
		\begin{supertabular}{R{0.3cm}C{0cm}L{7cm}}
			$D1$	& $\equiv$ & $\ensuremath{\exists}  hasInstrument.(\exists hasIT.Calliper \ensuremath{\sqcap}$ \\
			& & 	$\ensuremath{\exists} hasRM.Numeric  \ensuremath{\sqcap}$ \\
			& & 	$ \ensuremath{\exists} hasMat.(Steel) ) \sqcap$ \\
			& &  $\exists hasMeasure.\top$\\
			$D2$ 	& $\equiv$ & $\ensuremath{\exists} hasInstrument.(Ruler  \ensuremath{\sqcap} \ensuremath{\exists} hasMat.Wood)$\\
			& & $ \ensuremath{\sqcap} \ensuremath{\exists} hasMeasure.( \ensuremath{\exists} hasUnit.Centimeter) $ \\
		\end{supertabular}
	\end{exe}
	
	\begin{defi}[Projection sur une composante]
		Soient une ${\cal EL}^{++}_{\setminus {\cal D}}$-CBox  CB, ${\cal C}_{E_1}$, ..., ${\cal C}_{E_n}$ les $n$ composantes de CB (données arbitrairement), avec $R_1$, ..., $R_n$ les rôles de composantes associés, et une ${\cal EL}^{++}_{\setminus {\cal D}}$-description de concept $O \equiv \exists R_1.C_1 \sqcap \ldots \sqcap \exists R_n.C_n $. La projection de $O$ sur ${\cal C}_{E_i}$ est la description de concept $C_i$, $\forall i \in \{1,\ldots,n\}$.
		On la note $O^{R_i}$.
	\end{defi}

	Au sein d'une demande ou d'une offre, selon la projection correspondante, les composantes sont considérées inexistantes ou existantes, selon la définition suivante.

	\begin{defi}[Composante existante et inexistante]
		Etant donnée une ${\cal EL}^{++}_{\setminus {\cal D}}$-CBox CB, ${\cal C}_{E}$ une composante donnée, $R_{{\cal C}_{E}}$ le rôle de composante associé, et $O$ une ${\cal EL}^{++}_{\setminus {\cal D}}$-description de concept. On dit que ${\cal C}_{E}$ est existante dans $O$ si $O^{R_{{\cal C}_{E}}} \neq \top$, et inexistante sinon,\\
	\end{defi}
	\vspace{-0.5cm}
	\subsection{Classement sémantique par composante}
	\label{semantic}

	Nous proposons maintenant de classer les offres selon leur proximité sémantique par rapport à la demande, pour chaque composante. On doit ici formaliser la notion de plus grande proximité entre une offre et une demande, pour une composante donnée. On s'inspire de la notion de meilleure couverture \cite{Bcov} pour définir la plus grande proximité entre une offre et une demande par la maximisation de l'information commune entre elles. On obtient cela en minimisant respectivement les informations de la demande absentes de l'offre d'un côté, et les informations de l'offre absentes de la demande de l'autre. Ce sont les notions de Rest et Miss, basées sur le calcul de l'information commune entre l'offre et la demande (raisonnement de plus petit subsumant commun ou least common subsumer \cite{LCSinEL}), et sur le calcul de l'information de l'une manquante dans l'autre (différence sémantique \cite{DifferenceTeege}). 
	
	\begin{exe}
		\label{excours4}
		Soient $D$ une demande et $O$ une offre :
		\begin{supertabular}{R{0.3cm}C{0cm}L{7cm}}
			$D$	& $\equiv$ & $\exists hasInstrument(\exists hasIT.Calliper \ensuremath{\sqcap}$\\
			& &		$ \exists hasMat.Steel) \ensuremath{\sqcap}$\\
			& &		$ \exists hasMeasure(\exists hasDim.length \ensuremath{\sqcap}$\\
			& &		$ \exists hasUnit.centimeter)$\\
			$O$ 	& $\equiv$ & $\exists hasInstrument(\exists hasIT.Calliper \ensuremath{\sqcap}$\\
			& &		$\exists  hasMat.Metal \ensuremath{\sqcap}\exists hasRM.Numeric ) \ensuremath{\sqcap}$\\
			& &		$ \exists hasMeasure(\exists hasDim.length) $ \\
		\end{supertabular}
		
		Intuitivement, pour la composante $hasInstrument$, on voudrait que le Rest qui est la partie de la demande non couverte par l'offre soit $ \exists hasMat.Steel $, et que le Miss qui est la partie de l'offre qui n'est pas demandée dans la demande soit $\exists hasRM.Numeric$. Pour la composante $hasMeasure$, on voudrait que le Rest soit $\exists hasUnit.centimeter$, et que le Miss soit $\top$ (c'est-à-dire qu'il n'y ait pas de Miss). 
	\end{exe}
	
	Pour déterminer le Rest (resp. le Miss),  \cite{Bcov} suggère d'oter à la demande (resp. à l'offre) les informations communes aux deux. On rappelle donc les définitions de least commun subsumer (LCS) et de différence sémantique permettant de réaliser ces opérations dans les logiques de description.
	
	\begin{defi}[Least Common Subsumer, lcs \cite{LCSinEL}]
		Soient $C$ et $D$ deux descriptions de concepts appartenant à la logique \ensuremath{\mathcal{L}}. La description de concept $E$ est un plus petit subsumant commun de $C$ et $D$ si et seulement si : (a) $C$ \ensuremath{\sqsubseteq} $E$ et $D$ \ensuremath{\sqsubseteq} $E$, et (b) $E$ est le concept le plus spécifique (le plus petit par rapport à la subsomption) à respecter (a).
	\end{defi}
	Quand le LCS de $C$ et $D$ existe, il est souvent unique et noté $LCS(C,D)$, ou $LCS_{CB}(C,D)$ s'il est calculé par rapport à une CBox CB. Dans ce travail, encore en cours, nous faisons l'hypothèse que le LCS existe et est unique dans ${\cal EL}^{++}_{\setminus {\cal D}}$. Cela reste à démontrer.
	\begin{exe}
		\label{excours5} En reprenant la CBox de l'exemple \ref{excours}, on définit les concepts $A$, $B$ et $C$ ainsi :\\
		$A$ $ \equiv $ Steel \ensuremath{\sqcap} Analogic ; $B$  $ \equiv $  Iron \ensuremath{\sqcap} Numeric ; $C$  $ \equiv $  Oak\\
		Ainsi, $LCS_{CB}$($A$, $B$) \ensuremath{\equiv} $Metal$ \ensuremath{\sqcap} $RM$ puisque l'on a dans la Cbox $Steel$ ~\ensuremath{\sqsubseteq} $Metal$ et $Iron$ ~\ensuremath{\sqsubseteq} $Metal$ et puisque $Analogic$ ~\ensuremath{\sqsubseteq} $RM$ et $Numeric$ ~\ensuremath{\sqsubseteq}$ RM$.
		
		De la même manière
		$LCS_{CB}$($A$, $C$) \ensuremath{\equiv} $Mat$ et $LCS_{CB}$($B$, $C$) \ensuremath{\equiv} $Material$ car dans la CBox, $Mat$ est le concept le plus précis subsumant à la fois $Oak$ et $Steel$ (resp. $Oak$ et $Iron$) (et $Analogic$ et $Numeric$ sont subsumés par $\top$.)
	\end{exe}
	
	\begin{defi}[Différence sémantique  \cite{DifferenceTeege}]
		Soient deux descriptions de concepts $C$ et $D$ appartenant à la logique \ensuremath{\mathcal{L}}, avec $C$ \ensuremath{\sqsubseteq} $D$. La différence sémantique $C \ominus D$ est définie par :  $max_{\sqsubseteq} \{E \in \ensuremath{\mathcal{L}} | E \ensuremath{\sqcap} D \ensuremath{\equiv} C\}$.
	\end{defi}
	La différence sémantique peut ne pas être unique \cite{Bcov}. Comme le LCS, nous faisons ici l'hypothèse que la différence sémantique est unique dans ${\cal EL}^{++}_{\setminus {\cal D}}$. Cela reste à démontrer. On définit maintenant les Rest et Miss.

	%
	%
	%
	%

	\begin{defi}[Rest et Miss]
		Soient CB une ${\cal EL}^{++}_{\setminus {\cal D}}$-CBox, et $O$ et $D$ deux ${\cal EL}^{++}_{\setminus {\cal D}}$-descriptions de concepts. Le Rest de $D$ par $O$ et le Miss de $D$ par $O$ sont notés $Rest_D(O)$ et $Miss_D(O)$ et sont définis ainsi :\\
		$Rest_D(O) \equiv D \ominus LCS_{CB}(D,O)$\\
		$Miss_D(O) \equiv O \ominus LCS_{CB}(D,O)$\\
	\end{defi}
\vspace{-0.7cm}
	\begin{exe} Soit CB la CBox de l'exemple \ref{excours}. Soient $D$ et $O$ les descriptions : $D \equiv Steel \ensuremath{\sqcap} Analogic$ et $O \equiv Metal \ensuremath{\sqcap} Numeric$. On a : 
		$LCS_{CB}(D,O) \equiv Metal \sqcap RM$\\
		$Rest_D(O) \equiv (Steel \ensuremath{\sqcap} Analogic)  \ominus (Metal \ensuremath{\sqcap} RM) \ensuremath{\equiv} Steel \ensuremath{\sqcap} Analogic \equiv D$\\
		$Miss_D(O)\ensuremath{\equiv} (Metal \ensuremath{\sqcap} Numeric)  \ominus   (Metal \ensuremath{\sqcap} RM) \ensuremath{\equiv} Numeric$\\
		Ainsi $O$ ne couvre aucune information de $D$ et ajoute des informations superflues (le mode de lecture $Numeric$).
		
	\end{exe}

	Nous pouvons maintenant proposer une méthode de classement des offres par rapport à une demande. Tout d'abord, on ne classe que les offres qui ont au moins une composante renseignée commune avec la demande, offres que l'on appellera recommandations.  Afin de profiter de la structure par composante des recommandations et de la demande, comme évoqué précédemment, on classe les recommandations par rapport à chaque composante de la CBox. Pour chaque composante, les recommandations les plus proches de la demande sont celles qui optimisent le Rest, puis en cas d'égalité celles qui optimisent le Miss. Optimiser le Rest en premier permet d'assurer que le plus possible d'informations données dans la demande soient présentes dans les meilleures recommandations. Optimiser le Miss sert à départager les recommandations qui seraient semblables du point de vue du Rest. L'optimisation des Rest et Miss va consister dans un premier temps à les maximiser par rapport à la subsomption (puisque plus un concept est grand par rapport à la subsomption, plus il est général et moins il contient d'information). Dans un second temps, là-encore pour départager les ex-aequo, on optimisera Rest et Miss en minimisant leur longueur syntaxique.
	\begin{defi}[Longueur syntaxique d'une description]
		Soit $D$ une ${\cal EL}^{++}_{\setminus {\cal D}}$-description de concept. La longueur syntaxique de $D$, notée $|D|$ est le nombre de concepts atomiques contenus dans $D$.
	\end{defi}

		\begin{figure}
		\hspace{-0.5cm}
		\includegraphics[width=8cm,height=6cm]{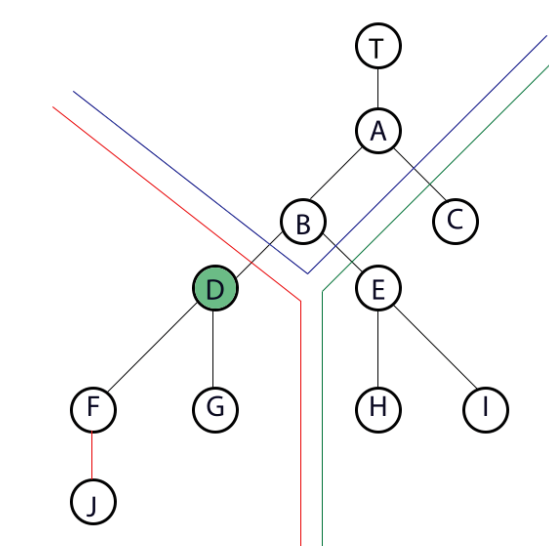}
		\caption{Exemple de positionnement des projections de recommandations ($A$, $B$, $C$, $E$, $F$, $G$, $H$, $I$ et $J$) et d'une demande ($D$) sur une composante. Les arêtes sont des liens de subsomption, et plus on monte, plus on est grand par rapport à la subsomption. Le concept en haut est $\top$.}
		\label{arbreproj}
		\vspace{-0.4cm}
	\end{figure}

		La figure \ref{arbreproj} illustre les 4 cas possibles qui peuvent survenir entre une demande $D$ et une recommandation, pour une composante. Dans la zone en rouge (en bas à gauche) se trouve la projection de $D$ sur la composante (rond en vert contenant $D$). $F$, $G$ et $J$ sont des projections possibles sur cette composante : ces projections sont subsumées par celle de $D$. On dit alors que la recommandation est plus précise que la demande. Dans la zone en bleu (en haut), on a les cas $A$, $B$ et $\top$ de projections de recommandation qui subsument celle de la demande. On dit alors que la recommandation est moins précise que la demande. Enfin dans la zone en vert (en bas à droite), $C$, $E$, $H$ et $I$ sont des cas où la recommandation est dite éloignée de la demande.

	L'algorithme \ref{Ordosem} permet de comparer les recommandations 2 à 2 par rapport à une demande, selon leur zone d'appartenance dans l'arbre (cf. figure \ref{arbreproj}). Il est facile de montrer que plus on monte dans cet arbre, plus le Rest est petit par rapport à la subsomption (i.e. moins il est bon), et plus on descend, plus le Miss est petit par rapport à la subsomption. Ainsi les recommandations plus précises maximisent le Rest (puisque ce dernier est alors $\top$) et sont donc meilleures que les recommandations moins précises, elles-mêmes meilleures que les recommandations éloignées. 
	
	
	A l'intérieur de chaque zone, on a les situations suivantes : \\
	- si deux recommandations sont plus précises alors on les départage en comparant leur miss : le miss le plus général (i.e. grand par rapport à la subsomption) est meilleur, et si les miss sont équivalents ou incomparables par rapport à la subsomption, le meilleur miss est le plus petit en taille . En cas de taille identique, les deux offres sont considérées équivalentes.\\
	- si deux recommandations sont moins précises alors on les départage en comparant leur rest : le rest le plus général par rapport à la subsomption est meilleur, et si les rest sont équivalents ou incomparables par rapport à la subsomption, le meilleur rest est le plus petit en taille. En cas de taille identique, les deux offres sont considérées équivalentes.\\
	- si deux recommandations sont éloignées, alors on cherche en premier celle qui maximise le rest par rapport à la subsomption, puis qui minimise le rest en taille, et en second si le rest n'a pu les départager, celle qui maximise le miss par rapport à la subsomption puis qui minimise le miss en taille. Si ces critères ne suffisent pas à départager les deux offres, alors elles sont considérées équivalentes.

\subsection{Approche multicritère}
\label{rankingMulti}

Notre objectif est maintenant de classer les recommandations par rapport à toutes les composantes, sachant qu'on a vu à la section précédente comment les classer deux à deux par rapport à chaque composante. 

 On voit facilement que l'on a à faire ici à un problème d'ordonnancement multicritère, où chaque composante est un critère. Nous proposons d'utiliser une relation de concordance relative \cite{Multicritere} pour le résoudre. C'est un des procédés les plus simples en ordonnancement multicritère. Nous rappelons maintenant son principe.
 
 Chaque recommandation $x$ est comparée avec toutes les autres recommandations $y$ sur chaque composante $i$, $1\leq i \leq n$, par la fonction $\phi_{i}(x,y)$ définie ci-dessous, qui nous donne le score relatif de $x$ par rapport $y$ pour $i$. 
 \vspace{-0.2cm}
 $$
 \phi_{i}(x,y) = \left\{
 \begin{array}{lll}
 si & x_{i} > y_{i}, & \phi_{i}(x,y)=1\\
 si & x_{i} = y_{i}, & \phi_{i}(x,y)=0\\
 si & x_{i} < y_{i}, & \phi_{i}(x,y)=-1\\
 \end{array}
 \right.
 $$
 On obtient alors le score relatif $c(x,y)$ de $x$ par rapport $y$ pour toutes les $n$ composantes en faisant la somme des $\phi_{i}(x,y)$ :
\vspace{-0.4cm}
 $$ c(x,y) =\sum_{i=0}^{n} \phi_{i}(x, y)$$
 On a bien entendu $c(x, y) = -c(y,x)$. Et on dira que $x$ est meilleure que $y$ ssi $c(x, y)$ ~\ensuremath{\geq} $c(y,x)$ (ou $c(x,y) ~\ensuremath{\geq} 0$).

Pour obtenir le score global de $x$, on fait ensuite la somme pour toutes les autres recommandations $y$ de $c(x,y)$ :
\vspace{-0.2cm}
$$ score(x) =\sum_{y, y\neq x} c(x,y) $$
Il suffit ensuite d'ordonner de façon décroissante les scores globaux pour classer les meilleures recommandations. Notons qu'il est possible de donner plus d'importance à des composantes $i$ en particulier en ajoutant un coefficient $v_i$ lors du calcul de ~\ensuremath{\phi_{i}}.

\begin{exe} On suppose que l'on a 3 recommandations $x$, $y$ et $z$, et 3 composantes $i$, $1\leq i \leq 3$. On suppose de plus que chaque recommandation possède une valeur pour chaque composante, ce qui permet de comparer les recommandations deux à deux pour chaque composante. Les valeurs de $x$ sont $(10, 5, 8)$, de $y$ sont $(11, 4, 8)$ et de $z$ sont $(9, 3, 7)$. 	
Ainsi, les scores relatifs des recommandations pour chaque composante sont :\\
$c(x, y)=\hspace{-0.1cm}\phi_{1}(10,11) + \phi_{2}(5,4) +\phi_{3}(8, 8) = -1 + 1 + 0 = 0$\\
$c(y, z)= ~\ensuremath{\phi_{1}}(11,9) + ~\ensuremath{\phi_{2}}(4,3) +~\ensuremath{\phi_{3}}(8, 7) = 1 + 1 + 1 = 3$\\
$c(x, z)= ~\ensuremath{\phi_{1}}(10,9) + ~\ensuremath{\phi_{2}}(5,3) +~\ensuremath{\phi_{3}}(8, 7) = 1 + 1 + 1 = 3$\\
Ainsi, $x$ et $y$ sont équivalentes, et $z$ est moins bonne, car $score(x)=score(y)=3$ et $score(z)=-6$.

Supposons maintenant que l'on veuille privilégier une composante en lui attribuant un coefficient de 3 (en laissant un coefficient de 1 aux deux autres). On a donc :\\
$c(x,y)\hspace{-0.1cm}=\hspace{-0.1cm}\phi_{1}(10,11)+\phi_{2}(5,4) +\phi_{3}(8,8)\hspace{-0.1cm}=\hspace{-0.1cm}-3+1+0=-2$\\
$c(y, z)= \ensuremath{\phi_{1}}(11,9) + ~\ensuremath{\phi_{2}}(4,3) +\ensuremath{\phi_{3}}(8, 7) = 3 + 1 + 1 = 5$\\
$c(x, z)= \ensuremath{\phi_{1}}(10,9) + \ensuremath{\phi_{2}}(5,3) +~\ensuremath{\phi_{3}}(8, 7) = 3 + 1 + 1 = 5$

Ainsi, $y$ est meilleure que $x$ et elles sont toutes deux meilleures que $z$, car $score(y) = 7$, $score(x)=3$ et $score(z)=-10$.\\
\end{exe}

\vspace{-0.5cm}
Il est facile d'appliquer le principe de concordance relative à notre cas d'étude, puisque l'algorithme \ref{Ordosem}~ implémente une fonction $\phi_{i}$, appelée ici $\phi_{R,\text{CB}}(D)(O_1,O_2)$ : les composantes $R$ de la CBox CB sont les composantes, $D$ est la demande et $O_1$ et $O_2$ sont les recommandations à comparer par rapport à $R$. Contrairement à l'exemple précédent, $\phi_{R,\text{CB}}(D)(O_1,O_2)$ compare $O_1$ et $O_2$ sur des critères sémantiques et non numériques.



\begin{algorithm}[htb]
		\small{
			\begin{algorithmic}[1]
				\REQUIRE Une ${\cal EL}^{++}_{\setminus {\cal D}}$-CBox CB avec ${\cal C}_{E_1}$, ..., ${\cal C}_{E_n}$ les $n$ composantes de CB (données arbitrairement) et $R_1$, ..., $R_n$ les rôles de composantes associés, une ${\cal EL}^{++}_{\setminus {\cal D}}$-description $D$  (la demande), et $m$ ${\cal EL}^{++}_{\setminus {\cal D}}$-descriptions $O_1,\ldots,O_m$ (les offres). 
				\ENSURE L'ensemble $e_O = \{(O_j, score_j)|~j\in\{1,..,m\}\}$ des couples composés de $O_j$ et du score associé $score_j$ par rapport à $D$, classés du plus grand score au plus petit.
				\STATE $e_O:= {\emptyset}$
				\STATE Initialisation de $score_1$ à $score_m$ à 0
				\FORALL{$(O_i, O_j) \in \{O_1,\ldots,O_m\}^2$ avec $j > i$}
					
					\FORALL{composante $R_k\in \{R_1,\ldots,R_n\}$} 
						\STATE $score_i := score_i+\phi_{R_k,\text{CB}}(D)(O_i,O_j)$ \COMMENT{~~~~//cf. algo. \ref{Ordosem}}
					\STATE $score_j:= score_j-\phi_{R_k,\text{CB}}(D)(O_i,O_j)$	
					\ENDFOR
					\ENDFOR
					\FORALL{$O_j$}
					\STATE $e_O$:=$e_O\cup (O_j,score_j)$ 
				\ENDFOR
				\STATE Tri des couples de $e_O$ par score décroissant.
				\STATE renvoyer $e_O$
			\end{algorithmic}
		}
	\caption{ Tri des recommandations}
	\label{AlgoQuiFaitTout}
\end{algorithm} 
	 \begin{algorithm*}[ht!]
	\begin{multicols}{2}
		\small{
			\begin{algorithmic}[1]
				\REQUIRE Une ${\cal EL}^{++}_{\setminus {\cal D}}$-CBox CB, un rôle de composante $R$ de CB, la description $D$ d'une demande, et les descriptions $O_1$ et $O_2$ de deux recommandations. 
				\ENSURE 1 si $O_1$ est meilleure que $O_2$ pour $D$ dans CB p/r à $R$, -1 si  $O_2$ est meilleure, et 0 si $O_1$ et $O_2$ sont équivalentes.\\
				
				\IF{$O_1^R \sqsubset D^R$}
				\IF{$O_2^R \sqsubset D^R$}
				\IF{$Miss_{D^R}(O_1^R) \sqsubset Miss_{D^R}(O_2^R)$}
				\STATE Renvoyer -1      	
				\ELSIF{$Miss_{D^R}(O_1^R) \sqsupset Miss_{D^R}(O_2^R)$}
				\STATE Renvoyer 1
				\ELSE
				\IF{$|Miss_{D^R}(O_1^R)| > |Miss_{D^R}(O_2^R)|$}
				\STATE Renvoyer -1      	
				\ELSIF{$|Miss_{D^R}(O_1^R)| < |Miss_{D^R}(O_2^R)|$}
				\STATE Renvoyer 1
				\ELSE
				\STATE Renvoyer 0
				\ENDIF
				\ENDIF
				\ELSIF{$O_2^R \equiv D^R$}
				\STATE Renvoyer -1
				\ELSE
				\STATE Renvoyer 1
				\ENDIF
				\ELSIF{$O_1^R \sqsupset D^R$}
				\IF{$O_2^R \sqsupset D^R$}
				\IF{$Rest_{D^R}(O_1^R) \sqsubset Rest_{D^R}(O_2^R)$}
				\STATE Renvoyer -1      	
				\ELSIF{$Rest_{D^R}(O_1^R) \sqsupset Rest_{D^R}(O_2^R)$}
				\STATE Renvoyer 1
				\ELSE
				\IF{$|Rest_{D^R}(O_1^R)| > |Rest_{D^R}(O_2^R)|$}
				\STATE Renvoyer -1      	
				\ELSIF{$|Rest_{D^R}(O_1^R)| < |Rest_{D^R}(O_2^R)|$}
				\STATE Renvoyer 1
				\ELSE
				\STATE Renvoyer 0
				\ENDIF
				\ENDIF
				\ELSIF{$O_2^R \sqsubseteq D^R$}
				\STATE Renvoyer -1
				\ELSE
				\STATE Renvoyer 1
				\ENDIF
				\ELSIF{$O_1^R \equiv D^R$}
				\IF{$O_2^R \equiv D^R$}
				\STATE Renvoyer 0
				\ELSE
				\STATE Renvoyer 1
				\ENDIF
				\ELSE
				\IF{($O_2^R \sqsubseteq D^R$) ou ($O_2^R \sqsupset D^R$)}
				\STATE Renvoyer -1
				\ELSE
				\IF{$Rest_{D^R}(O_1^R) \sqsubset Rest_{D^R}(O_2^R)$}
				\STATE Renvoyer -1      	
				\ELSIF{$Rest_{D^R}(O_1^R) \sqsupset Rest_{D^R}(O_2^R)$}
				\STATE Renvoyer 1
				\ELSE
				\IF{$|Rest_{D^R}(O_1^R)| > |Rest_{D^R}(O_2^R)|$}
				\STATE Renvoyer -1      	
				\ELSIF{$|Rest_{D^R}(O_1^R)| < |Rest_{D^R}(O_2^R)|$}
				\STATE Renvoyer 1
				\ELSE
				\IF{$Miss_{D^R}(O_1^R) \sqsubset Miss_{D^R}(O_2^R)$}
				\STATE Renvoyer -1      	
				\ELSIF{$Miss_{D^R}(O_1^R) \sqsupset Miss_{D^R}(O_2^R)$}
				\STATE Renvoyer 1
				\ELSE
				\IF{$|Miss_{D^R}(O_1^R)| > |Miss_{D^R}(O_2^R)|$}
				\STATE Renvoyer -1      	
				\ELSIF{$|Miss_{D^R}(O_1^R)|<|Miss_{D^R}(O_2^R)|$}
				\STATE Renvoyer 1
				\ELSE
				\STATE Renvoyer 0
				\ENDIF
				\ENDIF    				 
				\ENDIF
				\ENDIF
				\ENDIF
				\ENDIF
			\end{algorithmic}
		}
	\end{multicols}
	\caption{ Fonction $\phi_{R,\text{CB}}(D)(O_1,O_2)$ d'ordonnancement sémantique de deux recommandations $O_1$ et $O_2$ pour une demande $D$ par rapport à une composante $R$ de l'ontologie CB.
	}
	\label{Ordosem}\vspace{-0.3cm}
\end{algorithm*}    

\begin{exe}[Demandes et offres]
	En prenant la CBox CB de l'exemple 1, soit $D$ la demande utilisateur, $O_1$, $O_2$, $O_3$, $O_4$, les offres de notre base de connaissance. Les rôles de composantes sont abrégés en $R_1$ pour hasInstrument et $R_2$ pour hasMeasure.\\
	$D \equiv \exists R_1.(\exists hasMat.Metal \sqcap \exists hasIT.Ruler \sqcap \exists hasRM.Analogic) \sqcap \exists R_2.(\exists hasUnit.Centimeter \sqcap \exists hasDim.\top)$\\
	$0_1 \equiv \exists R_1.(\exists hasMat.Steel \sqcap \exists hasIT.Ruler \sqcap \exists hasRM.Analogic) \sqcap \exists R_2.(\exists hasUnit.Centimeter \sqcap \exists hasDim.\top)$\\
	$0_2 \equiv \exists R_1.(\exists hasMat.oak \sqcap \exists hasIT.Ruler \sqcap \exists hasRM.Analogic) \sqcap \exists R_2.(\exists hasUnit.Centimeter \sqcap \exists hasDim.\top)$\\
	$0_3 \equiv \exists R_1.(\exists hasMat.Metal \sqcap \exists hasIT.Ruler \sqcap \exists hasRM.Analogic) \sqcap \exists R_2.(\exists hasUnit.\top \sqcap \exists hasDim.\top)$\\
    $0_4 \equiv \exists R_1.(\exists hasMat.Wood \sqcap \exists hasIT.Ruler \sqcap \exists hasRM.\top) \sqcap \exists R_2.(\exists hasUnit.Centimeter \sqcap \exists hasDim.\top)$

	Pour chaque couple d'offres, on compare leurs composantes grâce à l'algorithme \ref{Ordosem}. Prenons quelques exemples, avec ($O_1$, $O_3$), ($O_1$, $O_2$) et ($O_2$, $O_4$) sur la composante $R_1$ avec $score_i$ le score calculé par l'algorithme:\\
	- $O_1^{R_1} \sqsubset D^{R_1}$ et $O_3^{R_1} \equiv D^{R_1}$, alors $score_1 := score_1-1$ et $score_3 := score_3+1$\\
	- $O_1^{R_1} \sqsubset D^{R_1}$ et $O_2^{R_1} \not\sqsubset D^{R_1}$ et $O_2^{R_1} \not\equiv D^{R_1}$, alors $score_1 := score_1+1$ et $score_2 := score_2-1$\\
	- $O_2^{R_1}$ et $O_4^{R_1}$ ne subsument et ne sont pas subsumées par  $D^{R_1}$, on compare donc leur Rest :
	
	$Rest_{D^{R_1}}(O_2^{R_1}) \equiv (\exists hasMat.Metal \sqcap \exists hasIT.Ruler \sqcap \exists hasRM.Analogic) \ominus (\exists hasMat.Oak \sqcap \exists hasIT.Ruler \sqcap \exists hasRM.Analogic) \equiv \exists hasMat.Metal$ \\
	$Rest_{D^{R_1}}(O_4^{R_1}) \equiv (\exists hasMat.Metal \sqcap \exists hasIT.Ruler \sqcap \exists hasRM.Analogic) \ominus (\exists hasMat.Wood \sqcap \exists hasIT.Ruler \sqcap \exists hasRM.\top )\equiv \exists hasMat.Metal \sqcap \exists hasRM.Analogic $\\
	$Rest_{D^{R_1}}(O_4^{R_1}) \sqsubset Rest_{D^{R_1}}(O_2^{R_1})$ alors $score_4 := score_4-1$ et $score_2 := score_2+1$
	
	On fait ainsi pour les deux composantes, chaque couple et on obtient les scores finaux :\\
	$score_1$ = 2, $score_2$ =0 , $score_3$ = 0, $score_4$ = -2 
	
	A l'issue de l'algorithme \ref{AlgoQuiFaitTout}, l'ensemble des offres avec leurs scores triées de façon décroissante est :\\
	{($O_1$, 2), ($O_2$, 0), ($O_3$, 0) ($O_4$, -2)}.
	
	$O_1$ est donc la recommandation la plus proche de $D$.
\end{exe}

\section{Travaux antérieurs}
\label{panorama}


Selon \cite{hacid}, les algorithmes de matchmaking sémantiques dépendent de deux caractéristiques : la présence ou non du calcul de la distance entre l'offre et la demande, et l'arité de la réponse. L'arité vaut 1-1 quand on classe plusieurs offres séparées, et elle est 1-N quand on classe plusieurs ensembles d'offres (pour mieux répondre à la demande). Dans \cite{SemanticMatchmaking} sont proposées des algorithmes de matchmaking basés sur les raisonnements d'abduction et de contraction de concept. L'abduction est ici utilisée  pour ajouter à une offre les descriptions de concept qu'il lui manque pour satisfaire une demande, tout en respectant les axiomes de la TBox correspondante. A contrario, la contraction de concept permet de retirer les descriptions de concept dans la demande dont la présence bloque la recommandation d'une offre. Les deux méthodes permettent de calculer une distance équivalente au nombre de modifications nécessaire pour passer de la demande à l'offre  et vis-versa, afin de classer les différentes offres de la plus proche à la plus éloignée. Elles sont toutes deux d'arités 1-1, c'est-à-dire qu'elle mette en relation une demande avec une seule offre. Le principe de meilleures couvertures\cite{Bcov} est différent, puisqu'il va mettre en relation une demande avec un ensemble d'offre dont la conjonction, appelée couverture, permet de répondre au maximum à la demande, il est donc d'arité 1-n. Le calcul de distance entre la demande et les ensembles d'offres utilise les concept de Rest et de Miss, qui correspondent respectivement aux caractéristiques que la couverture ne couvre pas dans la demande et les caractéristiques rajoutées par les réponses qui ne se trouvaient pas dans la demande.
Notre méthode de recommandation sémantique propose de reprendre le principe de Rest et de Miss pour le calcul des distances mais de l'appliquer à un système d'arité 1-1. On remplace un matchmaking d'arité 1-N par un matchmaking d'arité 1-1 mais sur plusieurs composantes. On passe alors d'une recherche combinatoire à un problème d'ordonnancement multicritère et nous espérons dans le futur prouver que cela fait baisser la complexité des algorithmes. Par ailleurs, les approches multicritères sont facilement personnalisables (avec par exemple l'ajout de coefficients pour pondérer les composantes) et donc adaptables à de nombreux contextes. 

\section{Conclusion}
\label{conclu}
\vspace{-0.2cm}
Dans le contexte du projet STAM dans le domaine de la métrologie, nous nous sommes intéressés au problème du classement de recommandations pour une requête utilisateur donnée. Nous avons formalisé la notion de proximité d'une recommandation par rapport à la requête : à partir des composantes, que nous avons définies comme les différentes parties de la requête et des recommandations, nous avons proposé un raisonnement dans la LD ${\cal EL}^+$ permettant de définir les recommandations retenues et de les comparer entre elles, pour obtenir un classement des recommandations pour chaque composante. Ce raisonnement s'appuie sur les raisonnements de différence et de LCS. Puis nous avons établi un ordonnancement global des recommandations selon une approche multicritère permettant d'intégrer les classements de toutes les composantes. 
Ce travail est encore en cours. Comme évoqué précédemment, il reste à démontrer l'existence d'un LCS et d'une différence sémantique unique pour deux ${\cal EL}^+$-descriptions par rapport à une ${\cal EL}^+$-CBox. De plus, l'étude de complexité théorique et calculatoire n'a pas été entamée.

\section*{Remerciements}
\vspace{-0.2cm}
Ce travail a été financé par le FEDER \mbox{https://ec.europa.eu/regional\_policy/fr/funding/erdf}.
\vspace{-0.27cm}
\bibliographystyle{plain}
\bibliography{./bibli}

\end{document}